\documentclass[aps,showpacs,prl,twocolumn,superscriptaddress]{revtex4-1}

\usepackage{graphicx}
\usepackage{dcolumn}
\usepackage{amsmath}
\usepackage{color}

\usepackage{subfigure}
\usepackage{setspace}
\usepackage{braket}
\usepackage{mathtools}

\begin{document}

\title{Correlation-driven charge and spin fluctuations in LaCoO$_3$}


\author{M. Karolak}
\affiliation{Institut f\"ur Theoretische Physik und Astrophysik, Universit\"at  W\"urzburg, Am Hubland, 97074 W\"urzburg, Germany}

\author{M. Izquierdo} 
\affiliation{European XFEL GmbH, Albert-Einstein-Ring 19, 22761 Hamburg, Germany}
\affiliation{I. Institut f\"ur Theoretische Physik, Universit\"at Hamburg, Jungiusstra{\ss}e 9, 20355 Hamburg, Germany}
\affiliation{Synchrotron Soleil, L'Orme des Merisiers St-Aubin, BP-48, 91192, Gif-sur-Yvette, France}

\author{S. L. Molodtsov}
\affiliation{European XFEL GmbH, Albert-Einstein-Ring 19, 22761 Hamburg, Germany}
\affiliation{Institute of Experimental Physics, Technische Universit\"at Bergakademie Freiberg, 09599 Freiberg, Germany}
\affiliation{ITMO University, Kronverkskiy pr. 49, 197101 St. Petersburg, Russia}

\author{A. I. Lichtenstein}
\affiliation{European XFEL GmbH, Albert-Einstein-Ring 19, 22761 Hamburg, Germany}
\affiliation{I. Institut f\"ur Theoretische Physik, Universit\"at Hamburg, Jungiusstra{\ss}e 9, 20355 Hamburg, Germany}

\date{\today}

\begin{abstract}
The spin transition in LaCoO$_3$ has been investigated within the density-functional theory + dynamical mean-field theory formalism using continuous time quantum Monte Carlo. Calculations on the experimental rhombohedral atomic structure with two Co sites per unit cell  show that an independent treatment of the Co atoms results in a ground state with strong charge fluctuations induced by electronic correlations. Each atom shows a contribution from either a $d^5$ or a $d^7$ state in addition to the main $d^6$ state. These states play a relevant role in the spin transition which can be understood as a low spin-high spin (LS-HS) transition with significant contributions ($\sim$ $10$\%) to the LS and HS states of $d^5$ and $d^7$ states respectively. A thermodynamic analysis reveals a  significant kinetic energy gain through introduction of charge fluctuations, which in addition to the potential energy reduction lowers the total energy of the system.
\end{abstract}


\maketitle


The interpretation of the remarkably low temperature ($\sim 100$K) spin transition in LaCoO$_3$ (LCO) has been an intriguing research topic for decades (see Ref. \cite{Ivanova09} for a review). A long standing debate regarding the origin of the transition is still ongoing with low-spin (LS, $t^6_{2g}e^0_g$) to high spin (HS, $t^4_{2g}e^2_g$) and low-spin to intermediate spin (IS, $t^5_{2g}e^1_g$) models with concomitant orbital ordering competing \cite{Heikes_1964,*Raccah67,korotin_1996}. The interest in this system has been recently boosted by the potential applications of this material and its doped phases in various optimized, environmental-friendly energy production domains \cite{Li08,*Alifanti04,*Kim10,*Buechler07,*Alvarez11}.
From the theoretical point LCO addresses one of the most challenging questions in solid state theory: the treatment of strongly correlated materials in which HS (IS) without long range magnetic order is stabilized \cite{Ivanova09}.
Although the addition of static local correlations to the density-functional theory (DFT+U) \cite{ldapu_review} improved the description of correlated materials \emph{with} long range magnetic order, the more advanced formalism combining DFT with dynamical mean-field theory (DMFT) \cite{ldadmft_review} has to be used to accurately treat systems in a paramagnetic state with local moments, like LCO \cite{zhang_2012,krapek_2012}. Many-body calculations on LCO have been performed using DFT+DMFT \cite{craco_2008} and the variational cluster approximation \cite{eder_2010}. In the first case, the contribution from LS, HS and IS was studied and the spin state transition described as a smooth crossover from the homogeneous LS state into a non-homogeneous mixture of all three spin states. On the other hand, the VCA calculations showed that only the LS and HS states have appreciable weight in the density matrix over a wide temperature range \cite{eder_2010}.
The most recent attempts to settle the spin transition in LCO by means of the DFT+DMFT approach \cite{zhang_2012,krapek_2012} used the numerically exact continuous-time quantum Monte Carlo (CT-QMC) \cite{ctqmc_review}. This methodology allows the inclusion of local dynamical effects and temperature, which is not possible within the inherently zero temperature DFT and DFT+U. In both cases, a LS-HS transition scenario results. No evidence of IS configuration was obtained, even upon going beyond the $d^6$ ionic picture \cite{krapek_2012}. Thus, including $d^7$ and $d^8$ charged states to describe the local dynamics of the system, has allowed to interpret the spin state transitions as an LS (with few HS ions) to LS-HS short range ordered phase, with a subsequent melting of the order at higher temperatures. At room temperature, when 50\% LS-HS population is expected, a Co(LS)-O-Co(HS) arrangement is anticipated. Most of the \textit{ab initio} many-body calculations discussed so far have been performed 
assuming equivalent Co atoms, with 
exception of a couple of DFT+U studies \cite{knizek_2006,knizek_2009}. 
In a groundbreaking study on a two band Hubbard model within DMFT Kune\v{s} and K\v{r}\'apek \cite{kunes_2011} showed that charge imbalance between sites can occur on purely electronic grounds and possibly consititues an important piece in the LCO spin state transition puzzle.

In this letter, we bolster this proposal by reporting on the first \textit{ab initio} study of correlation driven charge and spin fluctuations in LCO within the DFT+DMFT formalism for inequivalent Co atoms. Our results  show that symmetry breaking of the Co sites creates a correlation-driven charge and spin fluctuation of purely electronic origin.


LaCoO$_3$ is a perovskite system showing a small tetragonal distortion of the CoO$_6$ octahedra that varies with temperature, as determined from neutron diffraction \cite{radaelli_2002}. To describe it we use a fully \textit{ab initio} approach starting from density-functional calculations of LCO using the experimentally determined structures for 5K, 300K and 650K. We use a rhombohedral unit cell containing two formula units. The VASP code \cite{kresse_hafner} with projector augmented wave basis sets (PAW) \cite{Bloechl_PAW,kresse_joubert} and the Perdew-Burke-Ernzerhof \cite{PBE} functional was employed for DFT calculations. A Wannier type construction using projected local orbitals as described in \cite{Amadon08,DFT++} was applied to construct a local low energy model, which contains the on-site energies (crystal field splitting) and hoppings extracted from DFT. We focus here on the $3d$ orbitals of Co exclusively, avoiding the possible ambiguities that arise from the use of $d+p$ 
orbital models in DFT+DMFT, see e.g. \cite{parragh_2013,nio_dc_paper}. 
Since the system is not perfectly cubic a straightforward Wannier construction produces a basis that retains some on-site mixing between the Co $t_{2g}$ and $e_g$ orbitals. This local $t_{2g}$-$e_g$ hybridization is mostly a consequence of this specific choice of orbital representation, therefore we have performed a unitary transformation after the Wannier projection to minimize it. The usual choices here are a rotation into the so-called ``crystal field basis'' or into a basis that renders the DFT occupancy matrix $\rho_{ij}=\langle \hat{c}^\dagger_{i} \hat{c}^{\phantom{\dagger}}_j\rangle$ 
diagonal on each atom, see e.g. Refs. \onlinecite{lechermann_wannier,pavarini_review}. The data presented were obtained using latter approach. We have, however, explored both approaches and found that they yield similar off-diagonal elements in the frequency dependent DFT Green's function $G^0(i\omega_n)$ and the choice has no qualitative effect on any conclusions drawn in what follows.

On the level of DFT we find a set of three orbitals very close in energy (two $e^{\pi}_g$ and one $a_{1g}$, that we will for brevity call $t_{2g}$) and two orbitals ($e_g$) higher in energy by about $\Delta\sim 1.5$--$1.65$eV.
We solve an effective multiorbital Hubbard model within DMFT using a hybridization expansion CT-QMC solver using density-density interaction terms. To treat the two Co atoms in the unit cell independently we employ the so-called inhomogeneous or real space DMFT \cite{real_space_dmft}, where we have to solve an Anderson Impurity model for each correlated atom $\alpha$ in the unit cell and the atoms effectively interact via their respective baths of conduction electrons. The generalised five band Hubbard model including the Coulomb interaction and a double counting correction then contains the following terms

\begin{equation}
\hat{H}=\hat{H}^{0}-\sum_{\alpha,m,\sigma}\mu^{\rm DC}_\alpha \hat{n}_{\alpha m,\sigma}+\frac{1}{2}\sum_{\substack{\alpha, i, j \\ \sigma,\sigma^\prime}}U^{\sigma \sigma^\prime}_{\alpha ij} \hat{n}_{\alpha i,\sigma}\hat{n}_{\alpha j,\sigma^\prime},
\label{hamiltonian}
\end{equation}
where $\hat{H}^{0}$ is the (Kohn-Sham) DFT Hamiltonian, $\hat{n}_{\alpha i,\sigma}=\hat{c}^{\dagger}_{\alpha i\sigma}\hat{c}^{\phantom{\dagger}}_{\alpha i\sigma}$ and $\hat{c}^{\dagger}_{\alpha i\sigma}$ is the creation operator of an electron on site $\alpha$ in Wannier state $i$ and spin $\sigma$. The double counting, $\propto \mu^{\rm DC}_\alpha$, amounts to a shift of the chemical potential for the Co $3d$ shell and is determined self consistently. To obtain the Coulomb interaction matrix we employ the parametrization via the Slater integrals \cite{SlaterBook} connected to the average direct and exchange couplings $U$ and $J$: $F^0=U$, $J=1/14 (F^2+F^4)$ and $F^4= 0.625F^2$. The aforementioned basis transformation was applied here as well.

\begin{figure}[t]
\includegraphics[width=0.49\textwidth]{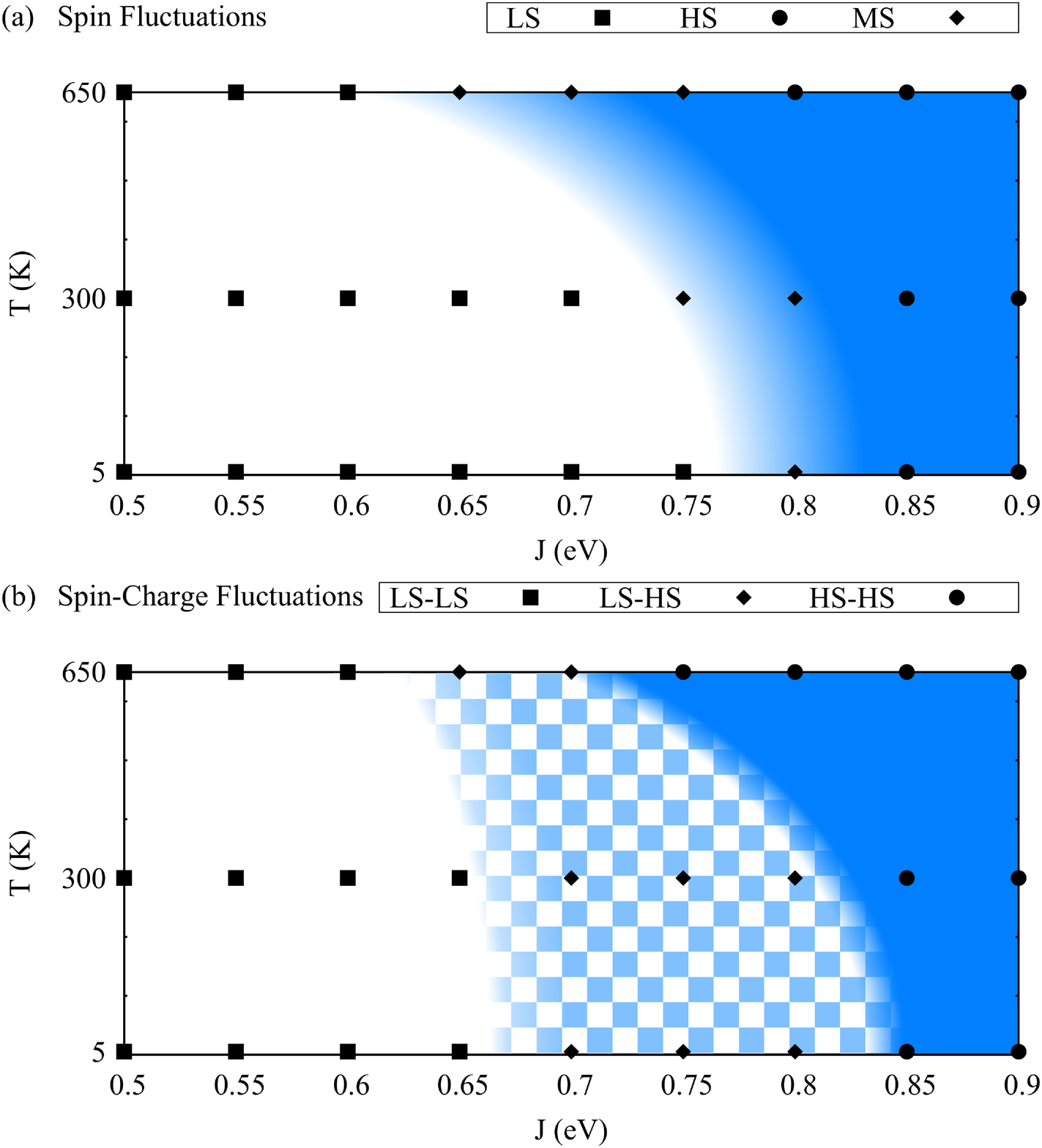}
\caption{(color online) Phase diagram as a function of the lattice structure temperature and the Hunds rule coupling $J$ for $U=3$eV. The colored (shaded) parts illustrate different regions in the phase diagram, white indicating LS and blue (gray) HS. (a) shows the homogeneous phase exhibiting the LS, MS and HS states as indicated by the data points. MS refers to a mixture of LS and HS states. (b) shows the data points and possible phase regions for the calculation with inequivalent Co atoms. The colored (shaded) areas are again only meant as an illustration, the checkerboard pattern now indicating the LS-HS phase where strong charge fluctuations are present.}
\label{phase_diagram}
\end{figure}

One important element to properly describe the system within DFT+DMFT is the appropriate choice of the Coulomb interaction. In the case of LCO, the problem is more delicate since after the LDA+U calculations that introduced the LS-IS model \cite{korotin_1996}, it has been widely believed that it is a strongly correlated electron system with an on-site Coulomb interaction of $U=8$eV. Recently, constrained DFT results proposed a value of $U=6$eV for the Co $3d$ shell in a $d+p$ model \cite{krapek_2012}.
Such high values for the Coulomb interaction might be appropriate for a $d+p$ model, are however too large for $d$ only calculations. Since we are not aware of any \textit{ab initio} estimates for the Co $3d$ shell only, measurements of the excitation gap in LCO have been used as a guide. A gap size of about $0.9$eV was measured from photoemission and absorption spectra \cite{abbate_1993}, while smaller values, $0.6$eV \cite{chainani_1992} and $0.3$eV \cite{arima_1993} were obtained from optical measurements. We were able to obtain a charge gap $\sim 1$ eV with a value of $U=3$eV, which will be used in all subsequent calculations. 

Since the value of the Hunds rule coupling $J$ is, for a fixed $U$ and crystal field splitting $\Delta$, the critical parameter for the spin state transition \cite{slater_ratio_2} we have calculated the system at different values of $J$. To account for the temperature, the calculations were performed for the experimental crystal lattice structures determined at the temperatures 5K, 300K and 650K \cite{radaelli_2002}. In the QMC solver we used the calculation temperatures 116K, 290K, 580K (equivalent to the inverse temperatures $\beta=100, 40, 20\text{eV}^{-1}$) for these structures respectively.

In a first approximation, the two Co atoms were constrained to be in the same charge and spin state. The calculations show that for the three crystal structures and their respective crystal fields a spin state transition occurs at about the point where $\Delta\sim 2 J$, i.e. for $J\sim 0.75-0.8$eV. In Fig. \ref{phase_diagram}a we have plotted our data for $U=3$eV as a function of the "lattice temperature" and $J$. We can see that there is a crossover region, that we call mixed spin (MS), between the homogeneous LS (white region) and HS (blue region) phases. The transition in the MS region is governed by an increased admixture of the HS contribution to the LS state. The hybridization expansion CT-QMC solver allows for an analysis of the local eigenstates contributing to the partition function during the imaginary time evolution. The states observed here are almost pure LS or HS ($> 90\%$ probability), i.e. $d^6_{S=0}$ and $d^6_{S=2}$ with no contribution from any IS ($d^6_{S=1}$) states.
Small ($\sim 3\%$) contributions of $d^5_{S=1/2}$ and $d^7_{S=3/2}$ are found in the LS and HS states respectively. 

\begingroup
\begin{table}[t]
\begin{ruledtabular}
\begin{tabular*}{0.45\textwidth}{@{\extracolsep{\fill}} lccccccc}
 $J$(eV) & Co (N$_{3d}$) & $d^6_{S=0}$ & $d^5_{1/2}$ & $d^7_{1/2}$ & $d^5_{3/2}$ &$d^7_{3/2}$ & $d^6_{2}$ \\
0.60 & 1 (6.0) & 93 & 4  & 3  & -   & - & -  \\
    & 2 (6.0) & 93 & 4  & 3  & -   & - & -  \\
\hline
0.75 & 1 (6.1) & 81 & 3 & 15  & -  & - &-  \\
    & 2 (5.9) &  - & -  & - & 12 & 3 & 82 \\
\hline
0.90 & 1 (6.0) &  - & - & -  & 3  & 7 & 87  \\
    & 2 (6.0) &  - & -  & - & 3 & 7 & 87 \\
\end{tabular*}
\end{ruledtabular}
\caption{Most probable many-body configurations for the 300K structure with two inequivalent Co atoms as a function of the Hunds coupling $J$ obtained from the analysis of the CT-QMC imaginary time evolution (in $\%$). Numbers missing to or sums exceeding $100\%$ are due to minor contributions of other atomic states and rounding. \label{tableConfig}}
\end{table}
\endgroup

\begin{figure}[t]
\centering
\includegraphics[width=0.49\textwidth]{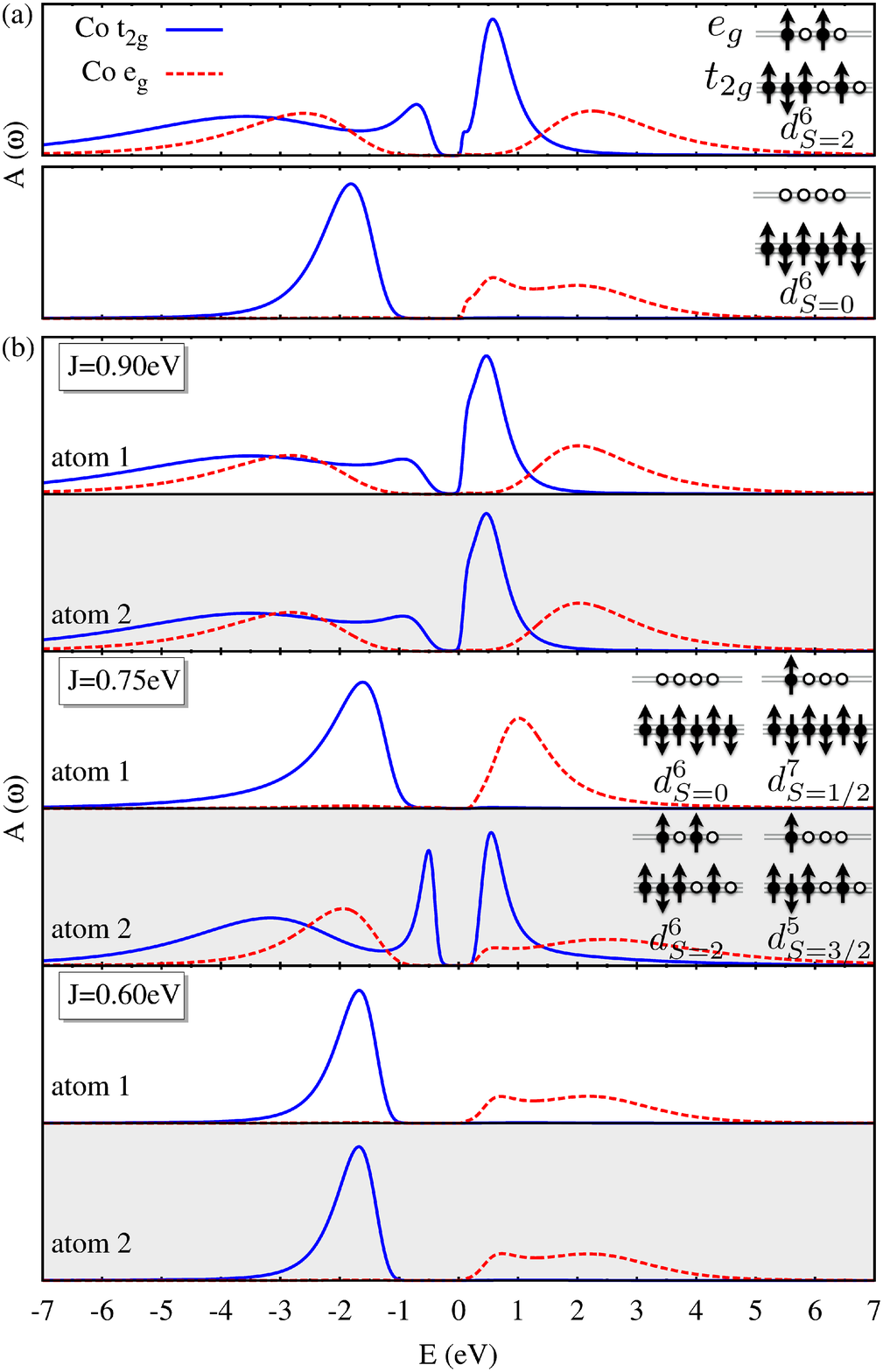}
\caption{(color online) Orbitally resolved spectral functions for LCO for the 300K crystal structure at solver temperature of 290K ($\beta=40$eV$^{-1}$). (a) The homogeneous HS ($J=0.9$eV, top) and LS spectra ($J=0.6$eV, bottom) with atomic states given as an inset. (b) Results in the asymmetric Co configuration for the values of $J$ indicated. The largest and second largest many-body contributions in the LS-HS ordered phase with $J=0.75$ eV are again given for a simplified octahedral crystal field.}
\label{lco_qmc_spec_1}
\end{figure}

In a second step, the constraint of equivalent Co atoms in the unit cell was removed. The calculations show that within this assumption, strong charge fluctuations between the two ions develop. This can happen spontaneously via noise introduced by the QMC procedure, but we have also introduced a small difference in the levels ($\mu^{\rm DC}_1$-$\mu^{\rm DC}_2=0.02$eV) in the first DMFT iteration to render the two atoms explicitly inequivalent \footnote{We want to note that a very large initial imbalance ($\mu^{\rm DC}_1$-$\mu^{\rm DC}_2\sim 1$eV) leads at large enough values of $J$ to a stabilization of a $d^5_{S=5/2}$ -- $d^7_{S=3/2}$ order.}. More concretely, a contribution of $d^5$ and $d^7$ states to the nominal $d^6$ average charge develops as a function of $J$. Thus, at small $J$ the two atoms both converge to the LS configuration as before, but with increasing $J$ charge fluctuations between the two atoms occur, see the LS-HS region in Fig. \ref{phase_diagram}b. Regarding 
the spin configuration in the LS-HS phase one atom will be in a predominantly LS and the other in a predominantly HS state. This is accompanied by the occupancies of the respective Co $3d$ shells to deviate from $N_{3d}=6.0$, see Tab. \ref{tableConfig}. Consequently, the QMC partition function shows sizeable contributions of $d^7_{S=1/2}$ on atom 1 and $d^5_{S=3/2}$ on atom 2, respectively. Other theoretical results and the interpretation of experimental data \cite{Raccah67,haverkort_2006} indicate that such a state exists in LCO at room temperature.
From Fig. \ref{phase_diagram} one can see that the spin transition can be studied as a function of the Hunds coupling $J$ and of the temperature. This implies, that the transition can be driven only by electronic means, as shown by previous model calculations \cite{Kunes11}. In the following, the detailed electronic structure configuration will be investigated as a function of $J$ assuming the 300K crystal structure, which exhibits all relevant features present in the whole temperature range.

The evolution of the Co $3d$ spectra as a function of $J$ is given in Fig. \ref{lco_qmc_spec_1}. The orbitally resolved spectral function (obtained via maximum entropy \cite{maxent} from the QMC Green's function) is shown for the homogeneous case for the LS and HS states of the 300K crystal structure in \ref{lco_qmc_spec_1}a. The strongest overall changes are visible for the unoccupied part of the spectra, suggesting that experimental techniques addressing those states will be relevant to understand the system in more detail \cite{Izquierdo_2014}. The low-spin state in Fig. \ref{lco_qmc_spec_1}a is closest to the DFT solution, the strongest modification is the rigid upward shift of the $e_g$ bands and, as a consequence, the gap opening between $t_{2g}$ and $e_g$ states. This is in accordance with combined DFT and cluster calculations \cite{abbate_1994} as well as recent QMC \cite{krapek_2012}. The formation of local moments in the higher temperature HS states leads to the 
appearance of 
additional features in the spectrum. As a result the gap changes its character from $t_{2g}-e_{g}$ to $t_{2g}-t_{2g}$ with incoherent $t_{2g}$ excitations on both gap edges. The occupied parts of the spectra exhibit a transfer of spectral weight away from the strong $t_{2g}$ excitation peak towards higher binding energies as the LS to HS crossover commences, see Ref. \cite{krapek_2012} and references therein. The spectral function for the asymmetric Co configuration is displayed in Fig. \ref{lco_qmc_spec_1}b for the LS-LS ($J=0.6$eV), the LS-HS ($J=0.75$eV) and the HS-HS ($J=0.9$eV) arrangements. Again, as the transition from LS-LS to HS-HS commences the first $t_{2g}$ excitation peak is largest for the LS-LS state, reduced in the LS-HS and even more so in the HS-HS state. Also, the progressive reduction of the gap expected from experimental studies is reproduced \cite{abbate_1994, haverkort_2006, Hu12}.

\begin{figure}[t]
\includegraphics[width=0.49\textwidth]{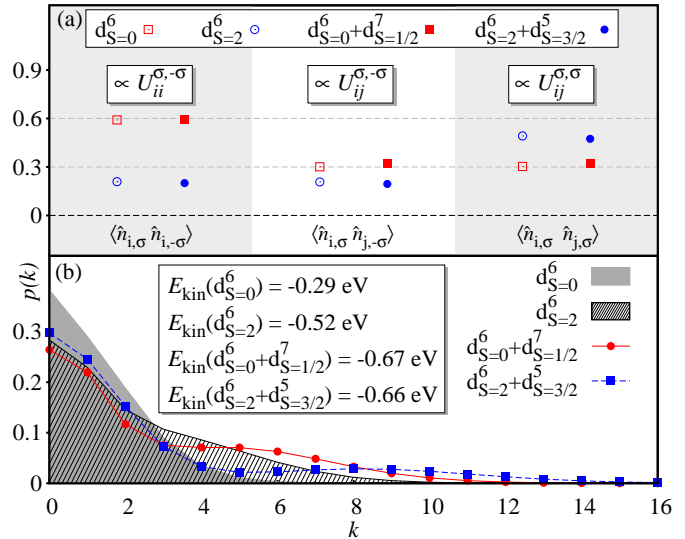}
\caption{(color online) Analysis of the total energy contributions for the 300K structure. 
(a) Orbitally averaged double occupancies in the homogeneous and inhomogeneous solutions (empty and solid symbols respectively) along with the corresponding Coulomb interaction term. The potential energy (obtained from the double occupancies) is lower in the LS-HS phase as compared to the homogeneous LS-LS phase, see text. (b) Histogram of the expansion order for 290K ($\beta=40$eV$^{-1}$) of the QMC diagrams contributing on one atom to the fermionic trace for the homogeneous (shaded) and the LS-HS phase (solid symbols). Its average is proportional to the kinetic energy. Numerical values correspond to the histograms shown.}
\label{energy}
\end{figure}

The tendency of the system to introduce charge fluctuations can be understood by analyzing the total energies within DMFT. The kinetic energy is accessible within QMC as $E_{\rm{kin}}=-T\langle k \rangle$, where $T$ is the temperature and $\langle k \rangle$ is the mean order of the QMC histogram, whereas the potential energy can be obtained from the double occupancy as $\frac{1}{2}\sum_{\alpha, i, j,\sigma,\sigma^\prime}U^{\sigma \sigma^\prime}_{\alpha ij} \langle \hat{n}_{\alpha i,\sigma}\hat{n}_{\alpha j,\sigma^\prime} \rangle$ \cite{held_1,leonov2011}.
As a consequence of the charge fluctuations the ``critical'' $J$ for the transition on one atom is lowered (via the $d^5$ admixture) allowing to introduce one HS site into the cell. The HS atom is then in a state of predominantly $d^6_{S=2}$ character with a contribution of a $d^5_{
S=3/2}$ state. The LS atom on the other hand is now in a $d^6_{S=0}$ with an admixture of $d^7_{S=1/2}$, which is higher in energy compared to the pure $d^6_{S=0}$ LS state. Balancing the energy loss of the LS atom with the energy gain brought about by the HS atom the mixed LS-HS arrangement constitutes a net gain in potential energy. The total energies can be analyzed in detail using Fig. \ref{energy}. The bulk of the potential energy gain can be understood as a reduction of the on-site double occupancies $\langle \hat{n}_{i,\sigma}\hat{n}_{i,-\sigma} \rangle$, see terms $\propto U_{ii}^{\sigma,-\sigma}$ in Fig. \ref{energy}a.  In the homogeneous LS state these terms are close to their maximum of $1$ for the $t_{2g}$ and zero for the $e_g$ states, thus giving the average value of $0.6$ per orbital, while dropping to about 0.2 per orbital in the HS state (empty symbols in Fig. \ref{energy}a). In the LS-LS state both atoms contribute with $\sim 0.6~U_{ii}$ per orbital, while in the LS-HS 
state one atom contributes $\sim 0.6~U_{ii}$, the other only $\sim 0.2~U_{ii}$ (solid symbols in Fig. \ref{energy}a). The net gain due to the contributions of the ``anti-Hund'' double occupancies $\propto U^{\sigma,-\sigma}_{ij}$ and the energy loss due to the ``Hund'' terms $\propto U^{\sigma,\sigma}_{ij}$ almost cancel. Using the proper numerical values for the double occupancies and interaction matrices we obtain the difference $\Delta E_{\rm pot}=E_{\rm pot}^{\rm LS-HS}-E_{\rm pot}^{\rm LS-LS}\approx -2.4$eV assuming $J=0.7$eV. At large enough $J$ the Hunds rule energy overcomes the crystal field and the HS-HS state becomes accessible.

The QMC histogram ($\sim -E_{\rm kin}$) shown in Fig. \ref{energy}b illustrates beautifully the appearance of charge fluctuations and can be interpreted using a reasoning similar to Ref. \cite{adriano_dp}. The QMC histogram for the pure LS and HS states shows the usual one peak structure, HS being more itinerant and exhibiting a shoulder in addition to the principal peak, while for the LS-HS state a clear two peak structure is observed. The first peak indicates standard processes that also occur for a single site, i.e. hopping from site $\alpha$ to site $\alpha$, while the second peak is of non-local origin and a consequence of the introduction of a second site $\alpha^\prime$. These processes have a larger kinetic energy (by absolute value) and thus are more itinerant than the homogeneous LS and HS phases. Concretely, both atoms in the LS-HS phase have a kinetic energy even larger (by absolute value) than in the HS homogeneous phase. This is a hallmark of the strong charge fluctuations between the two sites 
and constitues an additional gain in energy not accounted for in a single site picture.

The proposed mechanism can easily explain the low transition temperature observed in LCO on purely electronic grounds, since already for the 5K structure at solver temperature of $116$K the LS-HS phase is firmly established. Nevertheless, the discrepancy between lattice temperature and solver temperature distorts the phase diagram. Besides, some of the approximations used might influence the results. In future work it would be interesting to investigate how additional terms in the Coulomb interaction (spin-flip and pair-hopping) influence the observed behaviour. In addition, the inclusion of the O $2p$ states into the interacting model with a Coulomb interaction on the $p$ states ($U_{pp}$) and between the $d$ and $p$ states ($U_{pd}$) would be desirable, but is at the moment too expensive and an approximation of the Coulomb interaction difficult. Another interesting point would be how the charge and spin fluctuations found here couple to phonons, see e.g. \cite{Bari1972}. Finally, a more complex order than 
the reported checkerboard phase might also build up.

In summary, ab initio DFT+DMFT calculations for the two atomic unit cell of LaCoO$_3$ show that upon treating the Co atoms independently strong charge fluctuations develop in the system. As a consequence, the spin transition can be understood as a transition from a homogeneous LS to a LS-HS ordered state with strong charge fluctuations and subsequently into a homogeneous HS state. This provides a novel understanding of the system in which the charge fluctuations are present in the system from first principles. Angle resolved photoemission experiments as a function of the temperature and time-resolved XAS studies with femtosecond resolution should allow to verify the proposed model.

\begin{acknowledgments}
The authors thank G. Sangiovanni and M. Veit for fruitful discussions. M. I. acknowledges BMBF proposal 05K12GU2 for financial support. Financial support by the Deutsche Forschungsgemeinschaft (DFG) through SFB 668 is gratefully acknowledged. M.K. acknowledges support from the DFG via FOR1162.
\end{acknowledgments}

\bibliography{lco_prl_long}

\begin{thebibliography}{47}%
\makeatletter
\providecommand \@ifxundefined [1]{%
 \@ifx{#1\undefined}
}%
\providecommand \@ifnum [1]{%
 \ifnum #1\expandafter \@firstoftwo
 \else \expandafter \@secondoftwo
 \fi
}%
\providecommand \@ifx [1]{%
 \ifx #1\expandafter \@firstoftwo
 \else \expandafter \@secondoftwo
 \fi
}%
\providecommand \natexlab [1]{#1}%
\providecommand \enquote  [1]{``#1''}%
\providecommand \bibnamefont  [1]{#1}%
\providecommand \bibfnamefont [1]{#1}%
\providecommand \citenamefont [1]{#1}%
\providecommand \href@noop [0]{\@secondoftwo}%
\providecommand \href [0]{\begingroup \@sanitize@url \@href}%
\providecommand \@href[1]{\@@startlink{#1}\@@href}%
\providecommand \@@href[1]{\endgroup#1\@@endlink}%
\providecommand \@sanitize@url [0]{\catcode `\\12\catcode `\$12\catcode
  `\&12\catcode `\#12\catcode `\^12\catcode `\_12\catcode `\%12\relax}%
\providecommand \@@startlink[1]{}%
\providecommand \@@endlink[0]{}%
\providecommand \url  [0]{\begingroup\@sanitize@url \@url }%
\providecommand \@url [1]{\endgroup\@href {#1}{\urlprefix }}%
\providecommand \urlprefix  [0]{URL }%
\providecommand \Eprint [0]{\href }%
\providecommand \doibase [0]{http://dx.doi.org/}%
\providecommand \selectlanguage [0]{\@gobble}%
\providecommand \bibinfo  [0]{\@secondoftwo}%
\providecommand \bibfield  [0]{\@secondoftwo}%
\providecommand \translation [1]{[#1]}%
\providecommand \BibitemOpen [0]{}%
\providecommand \bibitemStop [0]{}%
\providecommand \bibitemNoStop [0]{.\EOS\space}%
\providecommand \EOS [0]{\spacefactor3000\relax}%
\providecommand \BibitemShut  [1]{\csname bibitem#1\endcsname}%
\let\auto@bib@innerbib\@empty
\bibitem [{\citenamefont {Ivanova}\ \emph {et~al.}(2009)\citenamefont
  {Ivanova}, \citenamefont {Ovchinnikov}, \citenamefont {Korshunov},
  \citenamefont {Eremin},\ and\ \citenamefont {Kazak}}]{Ivanova09}%
  \BibitemOpen
  \bibfield  {author} {\bibinfo {author} {\bibfnamefont {N.~B.}\ \bibnamefont
  {Ivanova}}, \bibinfo {author} {\bibfnamefont {S.~G.}\ \bibnamefont
  {Ovchinnikov}}, \bibinfo {author} {\bibfnamefont {M.~M.}\ \bibnamefont
  {Korshunov}}, \bibinfo {author} {\bibfnamefont {I.~M.}\ \bibnamefont
  {Eremin}}, \ and\ \bibinfo {author} {\bibfnamefont {N.~V.}\ \bibnamefont
  {Kazak}},\ }\href@noop {} {\bibfield  {journal} {\bibinfo  {journal} {Uspekhi
  Fizicheskikh Nauk}\ }\textbf {\bibinfo {volume} {179}},\ \bibinfo {pages}
  {837} (\bibinfo {year} {2009})}\BibitemShut {NoStop}%
\bibitem [{\citenamefont {Heikes}\ \emph {et~al.}(1964)\citenamefont {Heikes},
  \citenamefont {Miller},\ and\ \citenamefont {Mazelsky}}]{Heikes_1964}%
  \BibitemOpen
  \bibfield  {author} {\bibinfo {author} {\bibfnamefont {R.}~\bibnamefont
  {Heikes}}, \bibinfo {author} {\bibfnamefont {R.}~\bibnamefont {Miller}}, \
  and\ \bibinfo {author} {\bibfnamefont {R.}~\bibnamefont {Mazelsky}},\
  }\href@noop {} {\bibfield  {journal} {\bibinfo  {journal} {Physica}\ }\textbf
  {\bibinfo {volume} {30}},\ \bibinfo {pages} {1600} (\bibinfo {year}
  {1964})}\BibitemShut {NoStop}%
\bibitem [{\citenamefont {Raccah}\ and\ \citenamefont
  {Goodenough}(1967)}]{Raccah67}%
  \BibitemOpen
  \bibfield  {author} {\bibinfo {author} {\bibfnamefont {P.~M.}\ \bibnamefont
  {Raccah}}\ and\ \bibinfo {author} {\bibfnamefont {J.~B.}\ \bibnamefont
  {Goodenough}},\ }\href {\doibase 10.1103/PhysRev.155.932} {\bibfield
  {journal} {\bibinfo  {journal} {Phys. Rev.}\ }\textbf {\bibinfo {volume}
  {155}},\ \bibinfo {pages} {932} (\bibinfo {year} {1967})}\BibitemShut
  {NoStop}%
\bibitem [{\citenamefont {Korotin}\ \emph {et~al.}(1996)\citenamefont
  {Korotin}, \citenamefont {Ezhov}, \citenamefont {Solovyev}, \citenamefont
  {Anisimov}, \citenamefont {Khomskii},\ and\ \citenamefont
  {Sawatzky}}]{korotin_1996}%
  \BibitemOpen
  \bibfield  {author} {\bibinfo {author} {\bibfnamefont {M.~A.}\ \bibnamefont
  {Korotin}}, \bibinfo {author} {\bibfnamefont {S.~Y.}\ \bibnamefont {Ezhov}},
  \bibinfo {author} {\bibfnamefont {I.~V.}\ \bibnamefont {Solovyev}}, \bibinfo
  {author} {\bibfnamefont {V.~I.}\ \bibnamefont {Anisimov}}, \bibinfo {author}
  {\bibfnamefont {D.~I.}\ \bibnamefont {Khomskii}}, \ and\ \bibinfo {author}
  {\bibfnamefont {G.~A.}\ \bibnamefont {Sawatzky}},\ }\href@noop {} {\bibfield
  {journal} {\bibinfo  {journal} {Phys. Rev. B}\ }\textbf {\bibinfo {volume}
  {54}},\ \bibinfo {pages} {5309} (\bibinfo {year} {1996})}\BibitemShut
  {NoStop}%
\bibitem [{\citenamefont {Li}\ \emph {et~al.}(2008)\citenamefont {Li},
  \citenamefont {Bor{\'e}ave}, \citenamefont {Deloume},\ and\ \citenamefont
  {Gaillard}}]{Li08}%
  \BibitemOpen
  \bibfield  {author} {\bibinfo {author} {\bibfnamefont {N.}~\bibnamefont
  {Li}}, \bibinfo {author} {\bibfnamefont {A.}~\bibnamefont {Bor{\'e}ave}},
  \bibinfo {author} {\bibfnamefont {J.-P.}\ \bibnamefont {Deloume}}, \ and\
  \bibinfo {author} {\bibfnamefont {F.}~\bibnamefont {Gaillard}},\ }\href@noop
  {} {\bibfield  {journal} {\bibinfo  {journal} {Solid State Ionics}\ }\textbf
  {\bibinfo {volume} {179}},\ \bibinfo {pages} {1396} (\bibinfo {year}
  {2008})}\BibitemShut {NoStop}%
\bibitem [{\citenamefont {Alifanti}\ \emph {et~al.}(2004)\citenamefont
  {Alifanti}, \citenamefont {Kirchnerova}, \citenamefont {Delmon},\ and\
  \citenamefont {Klvana}}]{Alifanti04}%
  \BibitemOpen
  \bibfield  {author} {\bibinfo {author} {\bibfnamefont {M.}~\bibnamefont
  {Alifanti}}, \bibinfo {author} {\bibfnamefont {J.}~\bibnamefont
  {Kirchnerova}}, \bibinfo {author} {\bibfnamefont {B.}~\bibnamefont {Delmon}},
  \ and\ \bibinfo {author} {\bibfnamefont {D.}~\bibnamefont {Klvana}},\
  }\href@noop {} {\bibfield  {journal} {\bibinfo  {journal} {Applied Catalysis
  A: General}\ }\textbf {\bibinfo {volume} {262}},\ \bibinfo {pages} {167}
  (\bibinfo {year} {2004})}\BibitemShut {NoStop}%
\bibitem [{\citenamefont {Kim}\ \emph {et~al.}(2010)\citenamefont {Kim},
  \citenamefont {Qi}, \citenamefont {Dahlberg},\ and\ \citenamefont
  {Li}}]{Kim10}%
  \BibitemOpen
  \bibfield  {author} {\bibinfo {author} {\bibfnamefont {C.~H.}\ \bibnamefont
  {Kim}}, \bibinfo {author} {\bibfnamefont {G.}~\bibnamefont {Qi}}, \bibinfo
  {author} {\bibfnamefont {K.}~\bibnamefont {Dahlberg}}, \ and\ \bibinfo
  {author} {\bibfnamefont {W.}~\bibnamefont {Li}},\ }\href@noop {} {\bibfield
  {journal} {\bibinfo  {journal} {Science}\ }\textbf {\bibinfo {volume}
  {327}},\ \bibinfo {pages} {1624} (\bibinfo {year} {2010})}\BibitemShut
  {NoStop}%
\bibitem [{\citenamefont {B{\"u}chler}\ \emph {et~al.}(2007)\citenamefont
  {B{\"u}chler}, \citenamefont {Serra}, \citenamefont {Meulenberg},
  \citenamefont {Sebold},\ and\ \citenamefont {Buchkremer}}]{Buechler07}%
  \BibitemOpen
  \bibfield  {author} {\bibinfo {author} {\bibfnamefont {O.}~\bibnamefont
  {B{\"u}chler}}, \bibinfo {author} {\bibfnamefont {J.}~\bibnamefont {Serra}},
  \bibinfo {author} {\bibfnamefont {W.}~\bibnamefont {Meulenberg}}, \bibinfo
  {author} {\bibfnamefont {D.}~\bibnamefont {Sebold}}, \ and\ \bibinfo {author}
  {\bibfnamefont {H.}~\bibnamefont {Buchkremer}},\ }\href@noop {} {\bibfield
  {journal} {\bibinfo  {journal} {Solid State Ionics}\ }\textbf {\bibinfo
  {volume} {178}},\ \bibinfo {pages} {91} (\bibinfo {year} {2007})}\BibitemShut
  {NoStop}%
\bibitem [{\citenamefont {{\'A}lvarez-Galv{\'a}n}\ \emph
  {et~al.}(2011)\citenamefont {{\'A}lvarez-Galv{\'a}n}, \citenamefont
  {Constantinou}, \citenamefont {Navarro}, \citenamefont {Villoria},
  \citenamefont {Fierro},\ and\ \citenamefont {Efstathiou}}]{Alvarez11}%
  \BibitemOpen
  \bibfield  {author} {\bibinfo {author} {\bibfnamefont {M.~C.}\ \bibnamefont
  {{\'A}lvarez-Galv{\'a}n}}, \bibinfo {author} {\bibfnamefont {D.~A.}\
  \bibnamefont {Constantinou}}, \bibinfo {author} {\bibfnamefont {R.~M.}\
  \bibnamefont {Navarro}}, \bibinfo {author} {\bibfnamefont {J.~A.}\
  \bibnamefont {Villoria}}, \bibinfo {author} {\bibfnamefont {J.~L.~G.}\
  \bibnamefont {Fierro}}, \ and\ \bibinfo {author} {\bibfnamefont {A.~M.}\
  \bibnamefont {Efstathiou}},\ }\href@noop {} {\bibfield  {journal} {\bibinfo
  {journal} {Applied Catalysis B: Environmental}\ }\textbf {\bibinfo {volume}
  {102}},\ \bibinfo {pages} {291} (\bibinfo {year} {2011})}\BibitemShut
  {NoStop}%
\bibitem [{\citenamefont {Anisimov}\ \emph {et~al.}(1997)\citenamefont
  {Anisimov}, \citenamefont {Aryasetiawan},\ and\ \citenamefont
  {Lichtenstein}}]{ldapu_review}%
  \BibitemOpen
  \bibfield  {author} {\bibinfo {author} {\bibfnamefont {V.~I.}\ \bibnamefont
  {Anisimov}}, \bibinfo {author} {\bibfnamefont {F.}~\bibnamefont
  {Aryasetiawan}}, \ and\ \bibinfo {author} {\bibfnamefont {A.~I.}\
  \bibnamefont {Lichtenstein}},\ }\href@noop {} {\bibfield  {journal} {\bibinfo
   {journal} {J. Phys.: Condens. Matter}\ }\textbf {\bibinfo {volume} {9}},\
  \bibinfo {pages} {767} (\bibinfo {year} {1997})}\BibitemShut {NoStop}%
\bibitem [{\citenamefont {Kotliar}\ \emph {et~al.}(2006)\citenamefont
  {Kotliar}, \citenamefont {Savrasov}, \citenamefont {Haule}, \citenamefont
  {Oudovenko}, \citenamefont {Parcollet},\ and\ \citenamefont
  {Marianetti}}]{ldadmft_review}%
  \BibitemOpen
  \bibfield  {author} {\bibinfo {author} {\bibfnamefont {G.}~\bibnamefont
  {Kotliar}}, \bibinfo {author} {\bibfnamefont {S.~Y.}\ \bibnamefont
  {Savrasov}}, \bibinfo {author} {\bibfnamefont {K.}~\bibnamefont {Haule}},
  \bibinfo {author} {\bibfnamefont {V.~S.}\ \bibnamefont {Oudovenko}}, \bibinfo
  {author} {\bibfnamefont {O.}~\bibnamefont {Parcollet}}, \ and\ \bibinfo
  {author} {\bibfnamefont {C.~A.}\ \bibnamefont {Marianetti}},\ }\href
  {\doibase 10.1103/RevModPhys.78.865} {\bibfield  {journal} {\bibinfo
  {journal} {Rev. Mod. Phys.}\ }\textbf {\bibinfo {volume} {78}},\ \bibinfo
  {pages} {865} (\bibinfo {year} {2006})}\BibitemShut {NoStop}%
\bibitem [{\citenamefont {Zhang}\ \emph {et~al.}(2012)\citenamefont {Zhang},
  \citenamefont {Gorelov}, \citenamefont {Koch},\ and\ \citenamefont
  {Pavarini}}]{zhang_2012}%
  \BibitemOpen
  \bibfield  {author} {\bibinfo {author} {\bibfnamefont {G.}~\bibnamefont
  {Zhang}}, \bibinfo {author} {\bibfnamefont {E.}~\bibnamefont {Gorelov}},
  \bibinfo {author} {\bibfnamefont {E.}~\bibnamefont {Koch}}, \ and\ \bibinfo
  {author} {\bibfnamefont {E.}~\bibnamefont {Pavarini}},\ }\href {\doibase
  10.1103/PhysRevB.86.184413} {\bibfield  {journal} {\bibinfo  {journal} {Phys.
  Rev. B}\ }\textbf {\bibinfo {volume} {86}},\ \bibinfo {pages} {184413}
  (\bibinfo {year} {2012})}\BibitemShut {NoStop}%
\bibitem [{\citenamefont {K\ifmmode~\check{r}\else \v{r}\fi{}\'apek}\ \emph
  {et~al.}(2012)\citenamefont {K\ifmmode~\check{r}\else \v{r}\fi{}\'apek},
  \citenamefont {Nov\'ak}, \citenamefont {Kune\ifmmode~\check{s}\else
  \v{s}\fi{}}, \citenamefont {Novoselov}, \citenamefont {Korotin},\ and\
  \citenamefont {Anisimov}}]{krapek_2012}%
  \BibitemOpen
  \bibfield  {author} {\bibinfo {author} {\bibfnamefont {V.}~\bibnamefont
  {K\ifmmode~\check{r}\else \v{r}\fi{}\'apek}}, \bibinfo {author}
  {\bibfnamefont {P.}~\bibnamefont {Nov\'ak}}, \bibinfo {author} {\bibfnamefont
  {J.}~\bibnamefont {Kune\ifmmode~\check{s}\else \v{s}\fi{}}}, \bibinfo
  {author} {\bibfnamefont {D.}~\bibnamefont {Novoselov}}, \bibinfo {author}
  {\bibfnamefont {D.~M.}\ \bibnamefont {Korotin}}, \ and\ \bibinfo {author}
  {\bibfnamefont {V.~I.}\ \bibnamefont {Anisimov}},\ }\href {\doibase
  10.1103/PhysRevB.86.195104} {\bibfield  {journal} {\bibinfo  {journal} {Phys.
  Rev. B}\ }\textbf {\bibinfo {volume} {86}},\ \bibinfo {pages} {195104}
  (\bibinfo {year} {2012})}\BibitemShut {NoStop}%
\bibitem [{\citenamefont {Craco}\ and\ \citenamefont
  {M\"uller-Hartmann}(2008)}]{craco_2008}%
  \BibitemOpen
  \bibfield  {author} {\bibinfo {author} {\bibfnamefont {L.}~\bibnamefont
  {Craco}}\ and\ \bibinfo {author} {\bibfnamefont {E.}~\bibnamefont
  {M\"uller-Hartmann}},\ }\href {\doibase 10.1103/PhysRevB.77.045130}
  {\bibfield  {journal} {\bibinfo  {journal} {Phys. Rev. B}\ }\textbf {\bibinfo
  {volume} {77}},\ \bibinfo {pages} {045130} (\bibinfo {year}
  {2008})}\BibitemShut {NoStop}%
\bibitem [{\citenamefont {Eder}(2010)}]{eder_2010}%
  \BibitemOpen
  \bibfield  {author} {\bibinfo {author} {\bibfnamefont {R.}~\bibnamefont
  {Eder}},\ }\href {\doibase 10.1103/PhysRevB.81.035101} {\bibfield  {journal}
  {\bibinfo  {journal} {Phys. Rev. B}\ }\textbf {\bibinfo {volume} {81}},\
  \bibinfo {pages} {035101} (\bibinfo {year} {2010})}\BibitemShut {NoStop}%
\bibitem [{\citenamefont {Gull}\ \emph {et~al.}(2011)\citenamefont {Gull},
  \citenamefont {Millis}, \citenamefont {Lichtenstein}, \citenamefont
  {Rubtsov}, \citenamefont {Troyer},\ and\ \citenamefont
  {Werner}}]{ctqmc_review}%
  \BibitemOpen
  \bibfield  {author} {\bibinfo {author} {\bibfnamefont {E.}~\bibnamefont
  {Gull}}, \bibinfo {author} {\bibfnamefont {A.~J.}\ \bibnamefont {Millis}},
  \bibinfo {author} {\bibfnamefont {A.~I.}\ \bibnamefont {Lichtenstein}},
  \bibinfo {author} {\bibfnamefont {A.~N.}\ \bibnamefont {Rubtsov}}, \bibinfo
  {author} {\bibfnamefont {M.}~\bibnamefont {Troyer}}, \ and\ \bibinfo {author}
  {\bibfnamefont {P.}~\bibnamefont {Werner}},\ }\href {\doibase
  10.1103/RevModPhys.83.349} {\bibfield  {journal} {\bibinfo  {journal} {Rev.
  Mod. Phys.}\ }\textbf {\bibinfo {volume} {83}},\ \bibinfo {pages} {349}
  (\bibinfo {year} {2011})}\BibitemShut {NoStop}%
\bibitem [{\citenamefont {Kn{\'\i}{\v{z}}ek}\ \emph {et~al.}(2006)\citenamefont
  {Kn{\'\i}{\v{z}}ek}, \citenamefont {Jir{\'a}k}, \citenamefont
  {Hejtm{\'a}nek},\ and\ \citenamefont {Nov{\'a}k}}]{knizek_2006}%
  \BibitemOpen
  \bibfield  {author} {\bibinfo {author} {\bibfnamefont {K.}~\bibnamefont
  {Kn{\'\i}{\v{z}}ek}}, \bibinfo {author} {\bibfnamefont {Z.}~\bibnamefont
  {Jir{\'a}k}}, \bibinfo {author} {\bibfnamefont {J.}~\bibnamefont
  {Hejtm{\'a}nek}}, \ and\ \bibinfo {author} {\bibfnamefont {P.}~\bibnamefont
  {Nov{\'a}k}},\ }\href@noop {} {\bibfield  {journal} {\bibinfo  {journal} {J.
  Phys.: Condens. Matter}\ }\textbf {\bibinfo {volume} {18}},\ \bibinfo {pages}
  {3285} (\bibinfo {year} {2006})}\BibitemShut {NoStop}%
\bibitem [{\citenamefont {{Kn\'\i\ifmmode \check{z}\else \v{z}\fi{}ek, K. and
  Jir\'ak, Z. and Hejtm\'anek, J. and Nov\'ak, P. and Ku,
  W.}}(2009)}]{knizek_2009}%
  \BibitemOpen
  \bibfield  {author} {\bibinfo {author} {\bibnamefont {{Kn\'\i\ifmmode
  \check{z}\else \v{z}\fi{}ek, K. and Jir\'ak, Z. and Hejtm\'anek, J. and
  Nov\'ak, P. and Ku, W.}}},\ }\href {\doibase 10.1103/PhysRevB.79.014430}
  {\bibfield  {journal} {\bibinfo  {journal} {Phys. Rev. B}\ }\textbf {\bibinfo
  {volume} {79}},\ \bibinfo {pages} {014430} (\bibinfo {year}
  {2009})}\BibitemShut {NoStop}%
\bibitem [{\citenamefont {{Kune\ifmmode \check{s}\else \v{s}\fi{}, J. and
  K\ifmmode \check{r}\else \v{r}\fi{}\'apek, V.}}(2011)}]{kunes_2011}%
  \BibitemOpen
  \bibfield  {author} {\bibinfo {author} {\bibnamefont {{Kune\ifmmode
  \check{s}\else \v{s}\fi{}, J. and K\ifmmode \check{r}\else \v{r}\fi{}\'apek,
  V.}}},\ }\href {\doibase 10.1103/PhysRevLett.106.256401} {\bibfield
  {journal} {\bibinfo  {journal} {Phys. Rev. Lett.}\ }\textbf {\bibinfo
  {volume} {106}},\ \bibinfo {pages} {256401} (\bibinfo {year}
  {2011})}\BibitemShut {NoStop}%
\bibitem [{\citenamefont {Radaelli}\ and\ \citenamefont
  {Cheong}(2002)}]{radaelli_2002}%
  \BibitemOpen
  \bibfield  {author} {\bibinfo {author} {\bibfnamefont {P.}~\bibnamefont
  {Radaelli}}\ and\ \bibinfo {author} {\bibfnamefont {S.-W.}\ \bibnamefont
  {Cheong}},\ }\href@noop {} {\bibfield  {journal} {\bibinfo  {journal} {Phys.
  Rev. B}\ }\textbf {\bibinfo {volume} {66}},\ \bibinfo {pages} {094408}
  (\bibinfo {year} {2002})}\BibitemShut {NoStop}%
\bibitem [{\citenamefont {Kresse}\ and\ \citenamefont
  {Hafner}(1994)}]{kresse_hafner}%
  \BibitemOpen
  \bibfield  {author} {\bibinfo {author} {\bibfnamefont {G.}~\bibnamefont
  {Kresse}}\ and\ \bibinfo {author} {\bibfnamefont {J.}~\bibnamefont
  {Hafner}},\ }\href@noop {} {\bibfield  {journal} {\bibinfo  {journal} {J.
  Phys.: Condens. Matter}\ }\textbf {\bibinfo {volume} {6}},\ \bibinfo {pages}
  {8245} (\bibinfo {year} {1994})}\BibitemShut {NoStop}%
\bibitem [{\citenamefont {Bl\"ochl}(1994)}]{Bloechl_PAW}%
  \BibitemOpen
  \bibfield  {author} {\bibinfo {author} {\bibfnamefont {P.~E.}\ \bibnamefont
  {Bl\"ochl}},\ }\href {\doibase 10.1103/PhysRevB.50.17953} {\bibfield
  {journal} {\bibinfo  {journal} {Phys. Rev. B}\ }\textbf {\bibinfo {volume}
  {50}},\ \bibinfo {pages} {17953} (\bibinfo {year} {1994})}\BibitemShut
  {NoStop}%
\bibitem [{\citenamefont {Kresse}\ and\ \citenamefont
  {Joubert}(1999)}]{kresse_joubert}%
  \BibitemOpen
  \bibfield  {author} {\bibinfo {author} {\bibfnamefont {G.}~\bibnamefont
  {Kresse}}\ and\ \bibinfo {author} {\bibfnamefont {D.}~\bibnamefont
  {Joubert}},\ }\href {\doibase 10.1103/PhysRevB.59.1758} {\bibfield  {journal}
  {\bibinfo  {journal} {Phys. Rev. B}\ }\textbf {\bibinfo {volume} {59}},\
  \bibinfo {pages} {1758} (\bibinfo {year} {1999})}\BibitemShut {NoStop}%
\bibitem [{\citenamefont {Perdew}\ \emph {et~al.}(1996)\citenamefont {Perdew},
  \citenamefont {Burke},\ and\ \citenamefont {Ernzerhof}}]{PBE}%
  \BibitemOpen
  \bibfield  {author} {\bibinfo {author} {\bibfnamefont {J.~P.}\ \bibnamefont
  {Perdew}}, \bibinfo {author} {\bibfnamefont {K.}~\bibnamefont {Burke}}, \
  and\ \bibinfo {author} {\bibfnamefont {M.}~\bibnamefont {Ernzerhof}},\ }\href
  {\doibase 10.1103/PhysRevLett.77.3865} {\bibfield  {journal} {\bibinfo
  {journal} {Phys. Rev. Lett.}\ }\textbf {\bibinfo {volume} {77}},\ \bibinfo
  {pages} {3865} (\bibinfo {year} {1996})}\BibitemShut {NoStop}%
\bibitem [{\citenamefont {Amadon}\ \emph {et~al.}(2008)\citenamefont {Amadon},
  \citenamefont {Lechermann}, \citenamefont {Georges}, \citenamefont {Jollet},
  \citenamefont {Wehling},\ and\ \citenamefont {Lichtenstein}}]{Amadon08}%
  \BibitemOpen
  \bibfield  {author} {\bibinfo {author} {\bibfnamefont {B.}~\bibnamefont
  {Amadon}}, \bibinfo {author} {\bibfnamefont {F.}~\bibnamefont {Lechermann}},
  \bibinfo {author} {\bibfnamefont {A.}~\bibnamefont {Georges}}, \bibinfo
  {author} {\bibfnamefont {F.}~\bibnamefont {Jollet}}, \bibinfo {author}
  {\bibfnamefont {T.~O.}\ \bibnamefont {Wehling}}, \ and\ \bibinfo {author}
  {\bibfnamefont {A.~I.}\ \bibnamefont {Lichtenstein}},\ }\href {\doibase
  10.1103/PhysRevB.77.205112} {\bibfield  {journal} {\bibinfo  {journal} {Phys.
  Rev. B}\ }\textbf {\bibinfo {volume} {77}},\ \bibinfo {eid} {205112}
  (\bibinfo {year} {2008})}\BibitemShut {NoStop}%
\bibitem [{\citenamefont {{Karolak}}\ \emph {et~al.}(2011)\citenamefont
  {{Karolak}}, \citenamefont {{Wehling}}, \citenamefont {{Lechermann}},\ and\
  \citenamefont {{Lichtenstein}}}]{DFT++}%
  \BibitemOpen
  \bibfield  {author} {\bibinfo {author} {\bibfnamefont {M.}~\bibnamefont
  {{Karolak}}}, \bibinfo {author} {\bibfnamefont {T.~O.}\ \bibnamefont
  {{Wehling}}}, \bibinfo {author} {\bibfnamefont {F.}~\bibnamefont
  {{Lechermann}}}, \ and\ \bibinfo {author} {\bibfnamefont {A.~I.}\
  \bibnamefont {{Lichtenstein}}},\ }\href@noop {} {\bibfield  {journal}
  {\bibinfo  {journal} {J. Phys.: Condens. Matter}\ }\textbf {\bibinfo {volume}
  {23}},\ \bibinfo {pages} {085601} (\bibinfo {year} {2011})}\BibitemShut
  {NoStop}%
\bibitem [{\citenamefont {Parragh}\ \emph {et~al.}(2013)\citenamefont
  {Parragh}, \citenamefont {Sangiovanni}, \citenamefont {Hansmann},
  \citenamefont {Hummel}, \citenamefont {Held},\ and\ \citenamefont
  {Toschi}}]{parragh_2013}%
  \BibitemOpen
  \bibfield  {author} {\bibinfo {author} {\bibfnamefont {N.}~\bibnamefont
  {Parragh}}, \bibinfo {author} {\bibfnamefont {G.}~\bibnamefont
  {Sangiovanni}}, \bibinfo {author} {\bibfnamefont {P.}~\bibnamefont
  {Hansmann}}, \bibinfo {author} {\bibfnamefont {S.}~\bibnamefont {Hummel}},
  \bibinfo {author} {\bibfnamefont {K.}~\bibnamefont {Held}}, \ and\ \bibinfo
  {author} {\bibfnamefont {A.}~\bibnamefont {Toschi}},\ }\href {\doibase
  10.1103/PhysRevB.88.195116} {\bibfield  {journal} {\bibinfo  {journal} {Phys.
  Rev. B}\ }\textbf {\bibinfo {volume} {88}},\ \bibinfo {pages} {195116}
  (\bibinfo {year} {2013})}\BibitemShut {NoStop}%
\bibitem [{\citenamefont {Karolak}\ \emph {et~al.}(2010)\citenamefont
  {Karolak}, \citenamefont {Ulm}, \citenamefont {Wehling}, \citenamefont
  {Mazurenko}, \citenamefont {Poteryaev},\ and\ \citenamefont
  {Lichtenstein}}]{nio_dc_paper}%
  \BibitemOpen
  \bibfield  {author} {\bibinfo {author} {\bibfnamefont {M.}~\bibnamefont
  {Karolak}}, \bibinfo {author} {\bibfnamefont {G.}~\bibnamefont {Ulm}},
  \bibinfo {author} {\bibfnamefont {T.}~\bibnamefont {Wehling}}, \bibinfo
  {author} {\bibfnamefont {V.}~\bibnamefont {Mazurenko}}, \bibinfo {author}
  {\bibfnamefont {A.}~\bibnamefont {Poteryaev}}, \ and\ \bibinfo {author}
  {\bibfnamefont {A.}~\bibnamefont {Lichtenstein}},\ }\href@noop {} {\bibfield
  {journal} {\bibinfo  {journal} {Journal of Electron Spectroscopy and Related
  Phenomena}\ }\textbf {\bibinfo {volume} {181}},\ \bibinfo {pages} {11 }
  (\bibinfo {year} {2010})}\BibitemShut {NoStop}%
\bibitem [{\citenamefont {Lechermann}\ \emph {et~al.}(2006)\citenamefont
  {Lechermann}, \citenamefont {Georges}, \citenamefont {Poteryaev},
  \citenamefont {Biermann}, \citenamefont {Posternak}, \citenamefont
  {Yamasaki},\ and\ \citenamefont {Andersen}}]{lechermann_wannier}%
  \BibitemOpen
  \bibfield  {author} {\bibinfo {author} {\bibfnamefont {F.}~\bibnamefont
  {Lechermann}}, \bibinfo {author} {\bibfnamefont {A.}~\bibnamefont {Georges}},
  \bibinfo {author} {\bibfnamefont {A.}~\bibnamefont {Poteryaev}}, \bibinfo
  {author} {\bibfnamefont {S.}~\bibnamefont {Biermann}}, \bibinfo {author}
  {\bibfnamefont {M.}~\bibnamefont {Posternak}}, \bibinfo {author}
  {\bibfnamefont {A.}~\bibnamefont {Yamasaki}}, \ and\ \bibinfo {author}
  {\bibfnamefont {O.~K.}\ \bibnamefont {Andersen}},\ }\href@noop {} {\bibfield
  {journal} {\bibinfo  {journal} {Phys. Rev. B}\ }\textbf {\bibinfo {volume}
  {74}},\ \bibinfo {eid} {125120} (\bibinfo {year} {2006})}\BibitemShut
  {NoStop}%
\bibitem [{\citenamefont {Pavarini}\ \emph {et~al.}(2005)\citenamefont
  {Pavarini}, \citenamefont {Yamasaki}, \citenamefont {Nuss},\ and\
  \citenamefont {Andersen}}]{pavarini_review}%
  \BibitemOpen
  \bibfield  {author} {\bibinfo {author} {\bibfnamefont {E.}~\bibnamefont
  {Pavarini}}, \bibinfo {author} {\bibfnamefont {A.}~\bibnamefont {Yamasaki}},
  \bibinfo {author} {\bibfnamefont {J.}~\bibnamefont {Nuss}}, \ and\ \bibinfo
  {author} {\bibfnamefont {O.~K.}\ \bibnamefont {Andersen}},\ }\href@noop {}
  {\bibfield  {journal} {\bibinfo  {journal} {New Journal of Physics}\ }\textbf
  {\bibinfo {volume} {7}},\ \bibinfo {pages} {188} (\bibinfo {year}
  {2005})}\BibitemShut {NoStop}%
\bibitem [{\citenamefont {Potthoff}\ and\ \citenamefont
  {Nolting}(1999)}]{real_space_dmft}%
  \BibitemOpen
  \bibfield  {author} {\bibinfo {author} {\bibfnamefont {M.}~\bibnamefont
  {Potthoff}}\ and\ \bibinfo {author} {\bibfnamefont {W.}~\bibnamefont
  {Nolting}},\ }\href {\doibase 10.1103/PhysRevB.59.2549} {\bibfield  {journal}
  {\bibinfo  {journal} {Phys. Rev. B}\ }\textbf {\bibinfo {volume} {59}},\
  \bibinfo {pages} {2549} (\bibinfo {year} {1999})}\BibitemShut {NoStop}%
\bibitem [{\citenamefont {Slater}(1960)}]{SlaterBook}%
  \BibitemOpen
  \bibfield  {author} {\bibinfo {author} {\bibfnamefont {J.~C.}\ \bibnamefont
  {Slater}},\ }\href@noop {} {\emph {\bibinfo {title} {{Quantum theroy of
  Atomic Structure, Vol 1}}}}\ (\bibinfo  {publisher} {McGraw-Hill},\ \bibinfo
  {year} {1960})\BibitemShut {NoStop}%
\bibitem [{\citenamefont {Abbate}\ \emph {et~al.}(1993)\citenamefont {Abbate}
  \emph {et~al.}}]{abbate_1993}%
  \BibitemOpen
  \bibfield  {author} {\bibinfo {author} {\bibfnamefont {M.}~\bibnamefont
  {Abbate}} \emph {et~al.},\ }\href {\doibase 10.1103/PhysRevB.47.16124}
  {\bibfield  {journal} {\bibinfo  {journal} {Phys. Rev. B}\ }\textbf {\bibinfo
  {volume} {47}},\ \bibinfo {pages} {16124} (\bibinfo {year}
  {1993})}\BibitemShut {NoStop}%
\bibitem [{\citenamefont {Chainani}\ \emph {et~al.}(1992)\citenamefont
  {Chainani}, \citenamefont {Mathew},\ and\ \citenamefont
  {Sarma}}]{chainani_1992}%
  \BibitemOpen
  \bibfield  {author} {\bibinfo {author} {\bibfnamefont {A.}~\bibnamefont
  {Chainani}}, \bibinfo {author} {\bibfnamefont {M.}~\bibnamefont {Mathew}}, \
  and\ \bibinfo {author} {\bibfnamefont {D.~D.}\ \bibnamefont {Sarma}},\ }\href
  {\doibase 10.1103/PhysRevB.46.9976} {\bibfield  {journal} {\bibinfo
  {journal} {Phys. Rev. B}\ }\textbf {\bibinfo {volume} {46}},\ \bibinfo
  {pages} {9976} (\bibinfo {year} {1992})}\BibitemShut {NoStop}%
\bibitem [{\citenamefont {Arima}\ \emph {et~al.}(1993)\citenamefont {Arima},
  \citenamefont {Tokura},\ and\ \citenamefont {Torrance}}]{arima_1993}%
  \BibitemOpen
  \bibfield  {author} {\bibinfo {author} {\bibfnamefont {T.}~\bibnamefont
  {Arima}}, \bibinfo {author} {\bibfnamefont {Y.}~\bibnamefont {Tokura}}, \
  and\ \bibinfo {author} {\bibfnamefont {J.~B.}\ \bibnamefont {Torrance}},\
  }\href {\doibase 10.1103/PhysRevB.48.17006} {\bibfield  {journal} {\bibinfo
  {journal} {Phys. Rev. B}\ }\textbf {\bibinfo {volume} {48}},\ \bibinfo
  {pages} {17006} (\bibinfo {year} {1993})}\BibitemShut {NoStop}%
\bibitem [{\citenamefont {Tanabe}\ and\ \citenamefont
  {Sugano}(1954)}]{slater_ratio_2}%
  \BibitemOpen
  \bibfield  {author} {\bibinfo {author} {\bibfnamefont {Y.}~\bibnamefont
  {Tanabe}}\ and\ \bibinfo {author} {\bibfnamefont {S.}~\bibnamefont
  {Sugano}},\ }\href {\doibase 10.1143/JPSJ.9.766} {\bibfield  {journal}
  {\bibinfo  {journal} {Journal of the Physical Society of Japan}\ }\textbf
  {\bibinfo {volume} {9}},\ \bibinfo {pages} {766} (\bibinfo {year}
  {1954})}\BibitemShut {NoStop}%
\bibitem [{Note1()}]{Note1}%
  \BibitemOpen
  \bibinfo {note} {We want to note that a very large initial imbalance ($\mu
  ^{\protect \rm DC}_1$-$\mu ^{\protect \rm DC}_2\sim 1$eV) leads at large
  enough values of $J$ to a stabilization of a $d^5_{S=5/2}$ -- $d^7_{S=3/2}$
  order.}\BibitemShut {Stop}%
\bibitem [{\citenamefont {Haverkort}\ \emph {et~al.}(2006)\citenamefont
  {Haverkort}, \citenamefont {Hu}, \citenamefont {Cezar}, \citenamefont
  {Burnus}, \citenamefont {Hartmann}, \citenamefont {Reuther}, \citenamefont
  {Zobel}, \citenamefont {Lorenz}, \citenamefont {Tanaka}, \citenamefont
  {Brookes}, \citenamefont {Hsieh}, \citenamefont {Lin}, \citenamefont {Chen},\
  and\ \citenamefont {Tjeng}}]{haverkort_2006}%
  \BibitemOpen
  \bibfield  {author} {\bibinfo {author} {\bibfnamefont {M.~W.}\ \bibnamefont
  {Haverkort}}, \bibinfo {author} {\bibfnamefont {Z.}~\bibnamefont {Hu}},
  \bibinfo {author} {\bibfnamefont {J.~C.}\ \bibnamefont {Cezar}}, \bibinfo
  {author} {\bibfnamefont {T.}~\bibnamefont {Burnus}}, \bibinfo {author}
  {\bibfnamefont {H.}~\bibnamefont {Hartmann}}, \bibinfo {author}
  {\bibfnamefont {M.}~\bibnamefont {Reuther}}, \bibinfo {author} {\bibfnamefont
  {C.}~\bibnamefont {Zobel}}, \bibinfo {author} {\bibfnamefont
  {T.}~\bibnamefont {Lorenz}}, \bibinfo {author} {\bibfnamefont
  {A.}~\bibnamefont {Tanaka}}, \bibinfo {author} {\bibfnamefont {N.~B.}\
  \bibnamefont {Brookes}}, \bibinfo {author} {\bibfnamefont {H.~H.}\
  \bibnamefont {Hsieh}}, \bibinfo {author} {\bibfnamefont {H.-J.}\ \bibnamefont
  {Lin}}, \bibinfo {author} {\bibfnamefont {C.~T.}\ \bibnamefont {Chen}}, \
  and\ \bibinfo {author} {\bibfnamefont {L.~H.}\ \bibnamefont {Tjeng}},\ }\href
  {\doibase 10.1103/PhysRevLett.97.176405} {\bibfield  {journal} {\bibinfo
  {journal} {Phys. Rev. Lett.}\ }\textbf {\bibinfo {volume} {97}},\ \bibinfo
  {pages} {176405} (\bibinfo {year} {2006})}\BibitemShut {NoStop}%
\bibitem [{\citenamefont {Kune\ifmmode~\check{s}\else \v{s}\fi{}}\ and\
  \citenamefont {K\ifmmode~\check{r}\else \v{r}\fi{}\'apek}(2011)}]{Kunes11}%
  \BibitemOpen
  \bibfield  {author} {\bibinfo {author} {\bibfnamefont {J.}~\bibnamefont
  {Kune\ifmmode~\check{s}\else \v{s}\fi{}}}\ and\ \bibinfo {author}
  {\bibfnamefont {V.}~\bibnamefont {K\ifmmode~\check{r}\else
  \v{r}\fi{}\'apek}},\ }\href {\doibase 10.1103/PhysRevLett.106.256401}
  {\bibfield  {journal} {\bibinfo  {journal} {Phys. Rev. Lett.}\ }\textbf
  {\bibinfo {volume} {106}},\ \bibinfo {pages} {256401} (\bibinfo {year}
  {2011})}\BibitemShut {NoStop}%
\bibitem [{\citenamefont {Jarrell}\ and\ \citenamefont
  {Gubernatis}(1996)}]{maxent}%
  \BibitemOpen
  \bibfield  {author} {\bibinfo {author} {\bibfnamefont {M.}~\bibnamefont
  {Jarrell}}\ and\ \bibinfo {author} {\bibfnamefont {J.~E.}\ \bibnamefont
  {Gubernatis}},\ }\href {\doibase DOI: 10.1016/0370-1573(95)00074-7}
  {\bibfield  {journal} {\bibinfo  {journal} {Physics Reports}\ }\textbf
  {\bibinfo {volume} {269}},\ \bibinfo {pages} {133 } (\bibinfo {year}
  {1996})}\BibitemShut {NoStop}%
\bibitem [{\citenamefont {Izquierdo}\ \emph {et~al.}(2014)\citenamefont
  {Izquierdo}, \citenamefont {Karolak}, \citenamefont {Pontius}, \citenamefont
  {Trabant}, \citenamefont {Sch\"ussler-Langheine}, \citenamefont {Holldack},
  \citenamefont {F\"olisch}, \citenamefont {Kummer}, \citenamefont
  {Prabhakaran}, \citenamefont {Boothroyd}, \citenamefont {Efimov},
  \citenamefont {Spiwek}, \citenamefont {Belozerov}, \citenamefont {Poteryaev},
  \citenamefont {Lichtenstein},\ and\ \citenamefont
  {Molodtsov}}]{Izquierdo_2014}%
  \BibitemOpen
  \bibfield  {author} {\bibinfo {author} {\bibfnamefont {M.}~\bibnamefont
  {Izquierdo}}, \bibinfo {author} {\bibfnamefont {M.}~\bibnamefont {Karolak}},
  \bibinfo {author} {\bibfnamefont {N.}~\bibnamefont {Pontius}}, \bibinfo
  {author} {\bibfnamefont {C.}~\bibnamefont {Trabant}}, \bibinfo {author}
  {\bibfnamefont {C.}~\bibnamefont {Sch\"ussler-Langheine}}, \bibinfo {author}
  {\bibfnamefont {K.}~\bibnamefont {Holldack}}, \bibinfo {author}
  {\bibfnamefont {A.}~\bibnamefont {F\"olisch}}, \bibinfo {author}
  {\bibfnamefont {K.}~\bibnamefont {Kummer}}, \bibinfo {author} {\bibfnamefont
  {D.}~\bibnamefont {Prabhakaran}}, \bibinfo {author} {\bibfnamefont {A.~T.}\
  \bibnamefont {Boothroyd}}, \bibinfo {author} {\bibfnamefont {V.}~\bibnamefont
  {Efimov}}, \bibinfo {author} {\bibfnamefont {M.}~\bibnamefont {Spiwek}},
  \bibinfo {author} {\bibfnamefont {A.}~\bibnamefont {Belozerov}}, \bibinfo
  {author} {\bibfnamefont {A.}~\bibnamefont {Poteryaev}}, \bibinfo {author}
  {\bibfnamefont {A.}~\bibnamefont {Lichtenstein}}, \ and\ \bibinfo {author}
  {\bibfnamefont {S.}~\bibnamefont {Molodtsov}},\ }\href@noop {} {\  (\bibinfo
  {year} {2014})},\ \bibinfo {note} {to appear in Phys. Rev. B}\BibitemShut
  {NoStop}%
\bibitem [{\citenamefont {Abbate}\ \emph {et~al.}(1994)\citenamefont {Abbate},
  \citenamefont {Potze}, \citenamefont {Sawatzky},\ and\ \citenamefont
  {Fujimori}}]{abbate_1994}%
  \BibitemOpen
  \bibfield  {author} {\bibinfo {author} {\bibfnamefont {M.}~\bibnamefont
  {Abbate}}, \bibinfo {author} {\bibfnamefont {R.}~\bibnamefont {Potze}},
  \bibinfo {author} {\bibfnamefont {G.~A.}\ \bibnamefont {Sawatzky}}, \ and\
  \bibinfo {author} {\bibfnamefont {A.}~\bibnamefont {Fujimori}},\ }\href
  {\doibase 10.1103/PhysRevB.49.7210} {\bibfield  {journal} {\bibinfo
  {journal} {Phys. Rev. B}\ }\textbf {\bibinfo {volume} {49}},\ \bibinfo
  {pages} {7210} (\bibinfo {year} {1994})}\BibitemShut {NoStop}%
\bibitem [{\citenamefont {Hu}\ \emph {et~al.}(2012)\citenamefont {Hu},
  \citenamefont {Wu}, \citenamefont {Koethe}, \citenamefont {Barilo},
  \citenamefont {Shiryaev}, \citenamefont {Bychkov}, \citenamefont {Sch{\"
  u}ssler-Langeheine}, \citenamefont {Lorenz}, \citenamefont {Tanaka},
  \citenamefont {Hsieh}, \citenamefont {Lin}, \citenamefont {Chen},
  \citenamefont {Brookes}, \citenamefont {Agrestini}, \citenamefont {Chin},
  \citenamefont {Rotter},\ and\ \citenamefont {Tjeng}}]{Hu12}%
  \BibitemOpen
  \bibfield  {author} {\bibinfo {author} {\bibfnamefont {Z.}~\bibnamefont
  {Hu}}, \bibinfo {author} {\bibfnamefont {H.}~\bibnamefont {Wu}}, \bibinfo
  {author} {\bibfnamefont {T.~C.}\ \bibnamefont {Koethe}}, \bibinfo {author}
  {\bibfnamefont {S.~N.}\ \bibnamefont {Barilo}}, \bibinfo {author}
  {\bibfnamefont {S.~V.}\ \bibnamefont {Shiryaev}}, \bibinfo {author}
  {\bibfnamefont {G.~L.}\ \bibnamefont {Bychkov}}, \bibinfo {author}
  {\bibfnamefont {C.}~\bibnamefont {Sch{\" u}ssler-Langeheine}}, \bibinfo
  {author} {\bibfnamefont {T.}~\bibnamefont {Lorenz}}, \bibinfo {author}
  {\bibfnamefont {A.}~\bibnamefont {Tanaka}}, \bibinfo {author} {\bibfnamefont
  {H.~H.}\ \bibnamefont {Hsieh}}, \bibinfo {author} {\bibfnamefont {H.-J.}\
  \bibnamefont {Lin}}, \bibinfo {author} {\bibfnamefont {C.~T.}\ \bibnamefont
  {Chen}}, \bibinfo {author} {\bibfnamefont {N.~B.}\ \bibnamefont {Brookes}},
  \bibinfo {author} {\bibfnamefont {S.}~\bibnamefont {Agrestini}}, \bibinfo
  {author} {\bibfnamefont {Y.~Y.}\ \bibnamefont {Chin}}, \bibinfo {author}
  {\bibfnamefont {M.}~\bibnamefont {Rotter}}, \ and\ \bibinfo {author}
  {\bibfnamefont {L.~H.}\ \bibnamefont {Tjeng}},\ }\href@noop {} {\bibfield
  {journal} {\bibinfo  {journal} {New Journal of Physics}\ }\textbf {\bibinfo
  {volume} {14}},\ \bibinfo {pages} {123025} (\bibinfo {year}
  {2012})}\BibitemShut {NoStop}%
\bibitem [{\citenamefont {Held}\ \emph {et~al.}(2001)\citenamefont {Held},
  \citenamefont {McMahan},\ and\ \citenamefont {Scalettar}}]{held_1}%
  \BibitemOpen
  \bibfield  {author} {\bibinfo {author} {\bibfnamefont {K.}~\bibnamefont
  {Held}}, \bibinfo {author} {\bibfnamefont {A.~K.}\ \bibnamefont {McMahan}}, \
  and\ \bibinfo {author} {\bibfnamefont {R.~T.}\ \bibnamefont {Scalettar}},\
  }\href {\doibase 10.1103/PhysRevLett.87.276404} {\bibfield  {journal}
  {\bibinfo  {journal} {Phys. Rev. Lett.}\ }\textbf {\bibinfo {volume} {87}},\
  \bibinfo {pages} {276404} (\bibinfo {year} {2001})}\BibitemShut {NoStop}%
\bibitem [{\citenamefont {Leonov}\ \emph {et~al.}(2011)\citenamefont {Leonov},
  \citenamefont {Poteryaev}, \citenamefont {Anisimov},\ and\ \citenamefont
  {Vollhardt}}]{leonov2011}%
  \BibitemOpen
  \bibfield  {author} {\bibinfo {author} {\bibfnamefont {I.}~\bibnamefont
  {Leonov}}, \bibinfo {author} {\bibfnamefont {A.~I.}\ \bibnamefont
  {Poteryaev}}, \bibinfo {author} {\bibfnamefont {V.~I.}\ \bibnamefont
  {Anisimov}}, \ and\ \bibinfo {author} {\bibfnamefont {D.}~\bibnamefont
  {Vollhardt}},\ }\href {\doibase 10.1103/PhysRevLett.106.106405} {\bibfield
  {journal} {\bibinfo  {journal} {Phys. Rev. Lett.}\ }\textbf {\bibinfo
  {volume} {106}},\ \bibinfo {pages} {106405} (\bibinfo {year}
  {2011})}\BibitemShut {NoStop}%
\bibitem [{\citenamefont {{Amaricci}}\ \emph {et~al.}(2013)\citenamefont
  {{Amaricci}}, \citenamefont {{Parragh}}, \citenamefont {{Capone}},\ and\
  \citenamefont {{Sangiovanni}}}]{adriano_dp}%
  \BibitemOpen
  \bibfield  {author} {\bibinfo {author} {\bibfnamefont {A.}~\bibnamefont
  {{Amaricci}}}, \bibinfo {author} {\bibfnamefont {N.}~\bibnamefont
  {{Parragh}}}, \bibinfo {author} {\bibfnamefont {M.}~\bibnamefont {{Capone}}},
  \ and\ \bibinfo {author} {\bibfnamefont {G.}~\bibnamefont {{Sangiovanni}}},\
  }\href@noop {} {\bibfield  {journal} {\bibinfo  {journal} {arXiv:1310.3043~}\
  } (\bibinfo {year} {2013})}\BibitemShut {NoStop}%
\bibitem [{\citenamefont {{Bari, Robert A. and Sivardi\`ere,
  J.}}(1972)}]{Bari1972}%
  \BibitemOpen
  \bibfield  {author} {\bibinfo {author} {\bibnamefont {{Bari, Robert A. and
  Sivardi\`ere, J.}}},\ }\href {\doibase 10.1103/PhysRevB.5.4466} {\bibfield
  {journal} {\bibinfo  {journal} {Phys. Rev. B}\ }\textbf {\bibinfo {volume}
  {5}},\ \bibinfo {pages} {4466} (\bibinfo {year} {1972})}\BibitemShut
  {NoStop}%
\end{thebibliography}%

\end{document}